\documentclass[journal]{IEEEtran}

\usepackage[pdftex]{graphicx}
\graphicspath{{figures/}}
\DeclareGraphicsExtensions{.pdf,.jpeg,.png,.svg}
\usepackage{cite}
\usepackage{amsmath}
\usepackage{array}

\usepackage{amsthm}
\usepackage{caption, subcaption}
\usepackage{algorithm, algpseudocode}
\usepackage{bbm}
\usepackage[normalem]{ulem}
\usepackage[colorlinks=true, allcolors=blue]{hyperref}
\usepackage{siunitx}
\usepackage{xparse}

\usepackage[acronym]{glossaries}
\glsdisablehyper
\newacronym{sota}{SOTA}{state-of-the-art}

\newacronym{pac}{PAC}{probably approximately correct}
\newacronym{rv}{RV}{random variable}
\newacronym{iid}{i.i.d.}{independent and identically distributed}

\newacronym{ipm}{IPM}{integral probability metric}

\newacronym{dt}{DT}{digital twin}
\newacronym{ndt}{NDT}{network digital twin}
\newacronym{pt}{PT}{physical twin}

\newacronym{ai}{AI}{artificial intelligence}
\newacronym{gd}{GD}{gradient descent}
\newacronym{erm}{ERM}{empirical risk minimization}
\newacronym{mlp}{MLP}{multi-layer perceptron}
\newacronym{vae}{VAE}{variational auto-encoder}
\newacronym{cdl}{CDL}{contrastive density learning}

\newacronym{p-erm}{P-ERM}{pseudo-empirical risk minimization}

\newacronym{rl}{RL}{reinforcement learning}
\newacronym{mdp}{MDP}{Markov decision process}
\newacronym{pg}{PG}{policy gradient}

\newacronym{mab}{MAB}{multi-armed bandit}
\newacronym{cmab}{CMAB}{contextual multi-armed bandit}
\newacronym{dm}{DM}{direct method}
\newacronym{ips}{IPS}{inverse propensity score}

\newacronym{3d}{3D}{three dimensional}
\newacronym{2d}{2D}{two dimensional}
\newacronym{rt}{RT}{ray tracer}

\newacronym{tx}{Tx}{transmitter}
\newacronym{rx}{Rx}{receiver}
\newacronym{bs}{BS}{base station}
\newacronym{los}{LoS}{line of sight}
\newacronym{nlos}{NLoS}{non-line of sight}
\newacronym{mimo}{MIMO}{massive input massive output}
\newacronym{cfr}{CFR}{channel frequency response}
\newacronym{toa}{ToA}{time of arrival}
\newacronym[longplural=angles of arrival]{aoa}{AoA}{angle of arrival}
\newacronym[longplural=angles of departure]{aod}{AoD}{angle of departure}
\newacronym{xpd}{XPD}{cross-polarization discriminant}
\newacronym{ckm}{CKM}{channel knowledge map}

\newacronym{cp}{CP}{conformal prediction}

\newacronym{ppi}{PPI}{prediction-powered inference}

\newacronym{dr}{DR}{doubly-robust}
\newacronym{tdr}{TDR}{tuned doubly-robust}
\newacronym{ctdr}{CTDR}{context-tuned doubly-robust}
\newacronym{cdr}{CDR}{context-aware doubly-robust}

\usepackage{utils/commands}

\begin{document}

\title{Context-Aware Doubly-Robust Semi-Supervised Learning}

\author{
    \IEEEauthorblockN{%
        Clement~Ruah,\IEEEmembership{~Graduate~Student~Member,~IEEE,}
        Houssem~Sifaou,\IEEEmembership{~Member,~IEEE,}\\
        Osvaldo~Simeone,\IEEEmembership{~Fellow,~IEEE,} and
        Bashir~Al-Hashimi\\
    }
    \IEEEauthorblockA{%
        CIIPS, Department of Engineering, King’s College London, London, UK%
    }
    \thanks{%
        C. Ruah, H. Sifaou, and O. Simeone are with King’s Communications, Learning \& Information Processing (KCLIP) Lab.
        The work of C. Ruah was supported by the Faculty of Natural, Mathematical, and Engineering Sciences at King's College London.
        The work of H. Sifaou and O. Simeone was partially supported by the European Union’s Horizon Europe project CENTRIC (101096379).
        O. Simeone was also supported by the Open Fellowships of the EPSRC (EP/W024101/1) and by the EPSRC project (EP/X011852/1).
    }
}

\IEEEtitleabstractindextext{%
\begin{abstract}
The widespread adoption of \gls{ai} in next-generation communication systems is challenged by the heterogeneity of traffic and network conditions, which call for the use of highly contextual, site-specific, data.
A promising solution is to rely not only on real-world data, but also on synthetic pseudo-data generated by a \gls{ndt}.
However, the effectiveness of this approach hinges on the accuracy of the \gls{ndt}, which can vary widely across different contexts.
To address this problem, this paper introduces \gls{cdr} learning, a novel semi-supervised scheme that adapts its reliance on the pseudo-data to the different levels of fidelity of the \gls{ndt} across contexts.
\Gls{cdr} is evaluated on the task of downlink beamforming where it outperforms previous state-of-the-art approaches, providing a $24\%$ loss decrease when compared to \gls{dr} semi-supervised learning in regimes with low labeled data availability.
\glsreset{ai}\glsreset{ndt}\glsreset{cdr}
\end{abstract}

}

\maketitle
\IEEEdisplaynontitleabstractindextext

\section{Introduction}

We are currently in the \emph{scaling era} of \gls{ai}, where progress is largely driven by simultaneous increases in both data availability and computational resources \cite{xiao2024rethinking}.
However, applications in engineering systems such as telecommunication networks are inherently constrained by the available computational resources and, furthermore, their data are highly \emph{contextual}. 
For example, the data logs for the physical layer of a telecommunication network are site- and time-specific \cite{kaltenberger2024driving}.

The scarcity of data relevant to a given \emph{context}—such as the site geometry and traffic conditions—can be mitigated through the use of \glsemphpl{ndt} \cite{khan2022digital, jiang2023digital, ruah2024calibrating, abouamer2024geometry, chen2024neuromorphic, sifaou2024semi, hou2025automatic}. 
In fact, an \gls{ndt} can be used as a \emph{teacher model} that, under a given context, generates \emph{pseudo-data} that retains statistical properties similar to those of the real-world system.
These pseudo-data can then be used to supplement the available real data, enhancing the training of \gls{ai} models.

In this paper, we focus on a common \emph{semi-supervised learning} framework in which the designer has access not only to labeled data but also to a pool of unlabeled data. 
Additionally, a \emph{teacher model} is available to assign \emph{pseudo-labels} to the unlabeled data.
The conventional approach, referred to as \glsemph{p-erm}, trains a target model by minimizing the empirical loss computed over both labeled and pseudo-labeled data \cite{jiao2024learning}.
Consequently, the generalization performance of \gls{p-erm} is heavily dependent on the accuracy of the teacher model.

\begin{figure}
    \centering
    \begin{subfigure}{\columnwidth}
        \centering
        \includegraphics[keepaspectratio=true, width=3.1in]{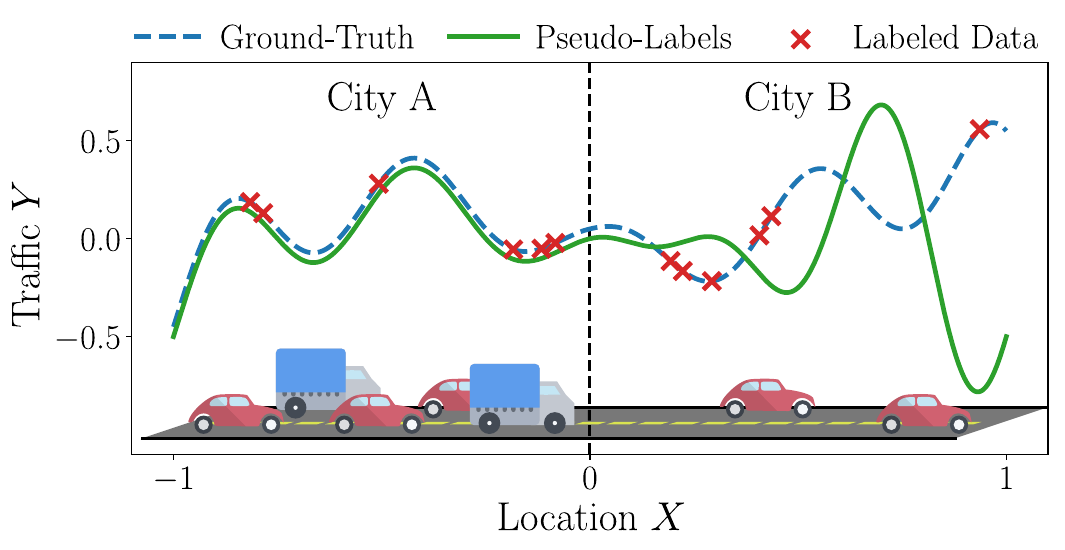}
    \end{subfigure}
    \begin{subfigure}{\columnwidth}
        \centering
        \includegraphics[keepaspectratio=true, width=3.1in]{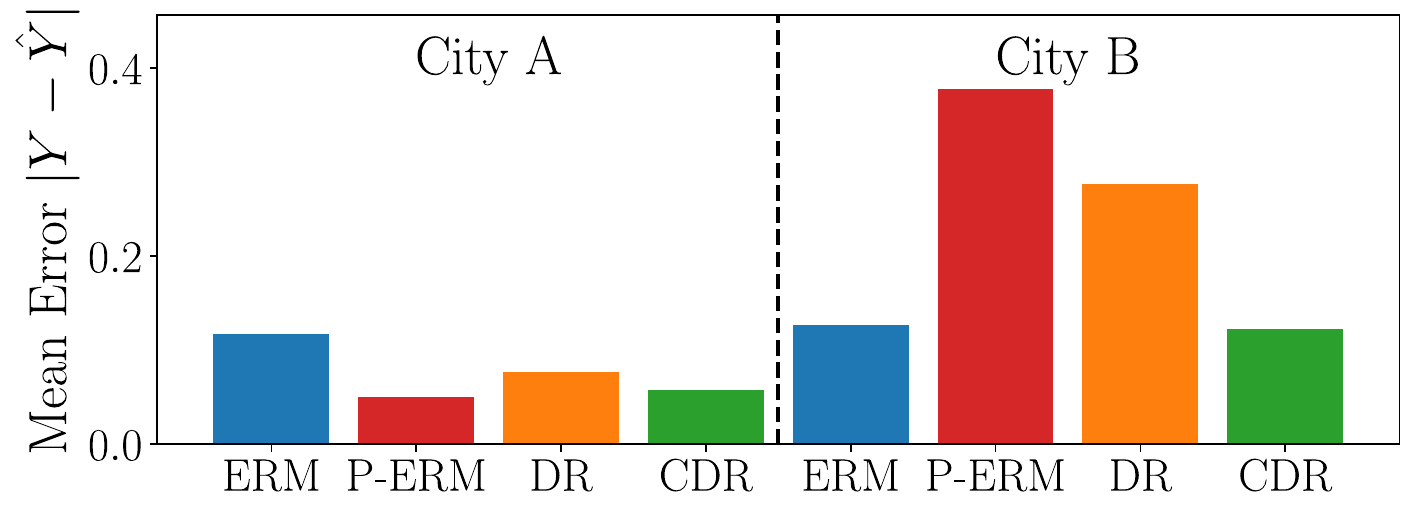}
    \end{subfigure}
    \caption{
    (top) Toy example dataset with scalar location covariate $X \in [-1, 1]$ and deterministic target $Y \in \R$ representing traffic conditions.
    (bottom) Average error $\abs{\hat{Y} - Y}$ per context $C \in \{\text{``City A''}, \text{``City B''}\}$ between ground-truth targets $Y$ and predictions $\hat{Y}$ from models trained using supervised \glsentryshort{erm} (blue), \glsentryshort{p-erm} (red), \glsentryshort{dr} (orange) and \glsentryshort{cdr} (green) learning.
    }
    \label{fig:toy_example}
\end{figure}

To address this limitation, reference \cite{zhu2024doubly} introduced \glsemph{dr} self-training, a methodology inspired by \cite{angelopoulos2023prediction}, which corrects for the bias introduced by inaccuracies in pseudo-labels during training.
\Gls{dr} was shown to outperform previous state-of-the-art semi-supervised learning algorithms such as FixMatch \cite{sohn2020fixmatch} and Mean Teacher \cite{tarvainen2017mean}, while displaying competitive performance with Meta Pseudo Labels \cite{pham2021meta} in classification settings \cite[Table~6]{zhu2024doubly}.

However, when applied to telecommunication networks, a key shortcoming of \gls{dr} is its failure to account for contextual information.
In practice, the accuracy of the teacher model can vary significantly depending on the context.
For instance, in the stylized scenario depicted in Fig.~\ref{fig:toy_example}, the pseudo-labels exhibit high accuracy inside the context $C=\text{``City A''}$ while they are significantly less reliable at locations within context $C=\text{``City B''}$.
When contextual information is available, an ideal training strategy should incorporate pseudo-labels in a manner that dynamically adjusts based on their reliability across different contexts. 

In this work, we introduce \glsemph{cdr} learning, an extension of \gls{dr} that adaptively adjusts the weighting of pseudo-labeled samples based on contextual information. 
As demonstrated in the toy example in Fig.~\ref{fig:toy_example}, \gls{cdr} effectively compensates for context-dependent errors in pseudo-labels, outperforming the state-of-the-art \gls{dr} method.

\section{Problem Definition} \label{sec:setting}

We consider a learning setting involving not only covariates $X \in \X$ and target variables, or labels, $Y \in \Y$, but also discrete \emph{context} variables $C \in \dset{K}$ associated to each pair $(X, Y)$.
The context variable $C$ \emph{stratifies} the input-output pairs $(X, Y)$ into $K \geq 1$ separate categories, representing properties shared by multiple data points.
These include, e.g., anagraphical features like age, or aspects associated with data collection, like location.

We focus on a \emph{semi-supervised} learning problem formulation characterized by a \emph{labeled dataset} $\D = \{(x_i, c_i, y_i)\}_{i=1}^{n} \iid P(X, C, Y)$, and an \emph{unlabeled dataset} $\tD = \{(\tx_i, \tc_i)\}_{i=1}^{N} \iid P(X, C)$.
Both datasets consist of \gls{iid} variables that follow the same population distribution $P(X, C, Y) = P(C) P(X|C) P(Y|X, C)$.
In addition to the datasets, learning can leverage a pre-trained \emph{teacher model} $f: \X \to \Y$ that assigns an estimate $f(X)$ of the target variable $Y \in \Y$ associated to covariate $X \in \X$ \cite{zhu2024doubly, amini2025self}.

Leveraging the datasets $\D$ and $\tD$, as well as the teacher model $f$, the goal is to train a parametric probabilistic model $p(y | x, \theta)$ by optimizing over the parameter vector $\theta \in \Theta$.
For a given loss function $\loss(\theta| x, y)$, the optimality of the trained parameter $\htheta$ is evaluated through the corresponding \emph{population loss} $\Loss(\htheta) = \E_{X, Y \sim P(X, Y)}[\loss(\htheta | X, Y)]$.
In what follows, we will use the notation
\begin{equation}
    L_{\loss, \D}(\theta) = \frac{1}{\abs{\D}} \sum_{(x, y) \in \D} \loss(\theta | x, y)
\end{equation}
for the empirical loss evaluated on a given dataset $\D$ and loss function $\loss(\theta| x, y)$.

\section{State of the Art} \label{sec:background}

The standard approach to address the semi-supervised learning problem outlined in the previous section is to augment the labeled dataset $\D$ with the synthetic labeled dataset $\tD^f = \{(\tx_i, f(\tx_i))\}_{i=1}^{N}$, in which the labels are imputed using the teacher model $f$, obtaining the \emph{pseudo-labels} $\{f(\tx_i)\}_{i=1}^{N}$ \cite{amini2025self}.
This leads to a training objective of the form
\begin{equation}
\label{eq:pseudo_erm_loss}
\begin{split}
L^{\pERM}(\theta) 
    &= L_{\loss, \D \cup \tD^f}(\theta), 
\end{split}
\end{equation}
whose optimization is referred as \glsemph{p-erm}.
The \gls{p-erm} training objective $L^{\pERM}(\theta)$ in \eqref{eq:pseudo_erm_loss} is a biased estimate of the population loss $\Loss(\theta)$, leading to an inconsistent estimate $\htheta^{\pERM} = \argmin_{\theta \in \Theta} L^{\pERM}(\theta)$ of the population loss minimizer $\theta^{*} = \argmin_{\theta \in \Theta} \Loss(\theta)$ \cite[Ch.~5]{van2000asymptotic}.

In order to circumvent this drawback, the \glsemph{dr} scheme introduced in \cite{zhu2024doubly} adopts the training objective
\begin{equation}
\label{eq:dr_loss}
L^{\DR}(\theta) =
    L_{\loss, \D^f \cup \tD^f}(\theta) -
    \underbrace{
        \left(
            L_{\loss, \D^f}(\theta) -
            L_{\loss, \D}(\theta)
        \right)
    }_{\text{bias correction}},
\end{equation}
where the dataset $\D^f = \{ (x_i, f(x_i)) \}_{i=1}^{n}$ replaces the available targets $\{y _i\}_{i=1}^{n}$ with their corresponding pseudo-labels $\{f(x_i)\}_{i=1}^{n}$.
In \eqref{eq:dr_loss}, the first term $L_{\loss, \D^f \cup \tD^f}(\theta)$ is the empirical loss based solely on pseudo-labels, while the second term corrects for the bias caused by the use of the teacher model.
This leads to an unbiased estimate of the population loss, i.e., $\E[L^{\DR}(\theta)] = \Loss(\theta)$, and to a consistent estimate $\htheta^{\DR} = \argmin_{\theta \in \Theta} L^{\DR}(\theta)$ of the population loss minimizer $\theta^{*}$.

\section{Context-Aware Doubly-Robust Self-training} \label{sec:cdr}

An important drawback of \gls{dr} is that it applies the same bias correction term to all data contexts irrespective of the quality of the pseudo-labels.
To address this limitation, this section introduces \glsemph{cdr} semi-supervised learning, a novel context-aware generalization of \gls{dr}.

\subsection{Context-Aware Doubly-Robust Training Objective}

\gls{cdr} starts by partitioning the labeled and unlabeled datasets as $\D_c = \{(x^c_i, y^c_i)\}_{i=1}^{n_c} = \{ (x_i, y_i) | 1 \leq i \leq n, c_i = c \}$ and $\tD_c = \{ \tx^c_i \}_{i=1}^{N_c} = \{ \tx_i | 1 \leq i \leq N, \tc_i = c \}$, respectively, for each value $c \in \dset{K}$ of the context variable $C$.
The datasets $\D^f$ and $\tD^f$ including the pseudo-labels are similarly partitioned as $\D^f_c = \{ (x^c_i, f(x^c_i)) \}_{i=1}^{n_c}$ and $\tD^f_c = \{ (\tx^c_i, f(\tx^c_i)) \}_{i=1}^{N_c}$ for $c \in \dset{K}$.
With these stratified datasets, the \gls{cdr} objective generalizes the \gls{dr} training loss in \eqref{eq:dr_loss} by breaking down the contributions of the labeled and unlabeled data across the values of the context variable $C$ and introducing a tuning parameter $\lambda_c \in [0, 1]$ for each context value $c$.
This yields the criterion
\begin{equation}
\label{eq:cdr_loss}
\begin{split}
L^{\CDR(\lambda)}_{\loss}(\theta) = 
    &\sum_{c=1}^{K} \bigg\{
        \frac{\lambda_c N_c}{N} L_{\loss, \tD^f_c}(\theta) \\
        & + \left(
            \frac{n_c}{n} L_{\loss, \D_c}(\theta) -
            \frac{\lambda_c n_c}{n} L_{\loss, \D^f_c}(\theta)
        \right)
    \bigg\},
\end{split}
\end{equation}
where $\lambda = [\lambda_1, ..., \lambda_K] \in [0, 1]^K$ is a \emph{tuning parameter vector}.

The inclusion of the parameters $\lambda_c$ in \eqref{eq:cdr_loss} allows \gls{cdr} to account for the extent to which the trained model relies on the pseudo-labeled data $\tD^f_c$ at context $c$.
Specifically, setting $\lambda_c = 0$ forces \gls{cdr} not to use the pseudo-labels for context $c$, while larger values of $\lambda_c$ ensure the use of the pseudo-labels for context $c$ in the estimate \eqref{eq:cdr_loss} \cite{angelopoulos2023ppi}.
Note that the \gls{dr} objective in \eqref{eq:dr_loss} is recovered as a special case of the \gls{cdr} objective $L^{\CDR(\lambda)}_{\loss}(\theta)$ when all entries of $\lambda$ are set $\lambda_c = 1 / (1 + n / N)$.
By varying the tuning parameter $\lambda_c$ with the context $c$, \gls{cdr} gains the flexibility to adjust the estimate \eqref{eq:cdr_loss} to different levels of accuracy of the teacher model, while maintaining unbiasedness, i.e., the equality $\E[L^{\CDR(\lambda)}_{\loss}(\theta)] = \Loss(\theta)$, regardless of the value of $\lambda$.

\begin{figure}
    \centering
    \begin{subfigure}{0.43\columnwidth}
        \centering
        \includegraphics[trim={0.1in 0 1.4in 0}, clip, keepaspectratio=true, width=1.45in]{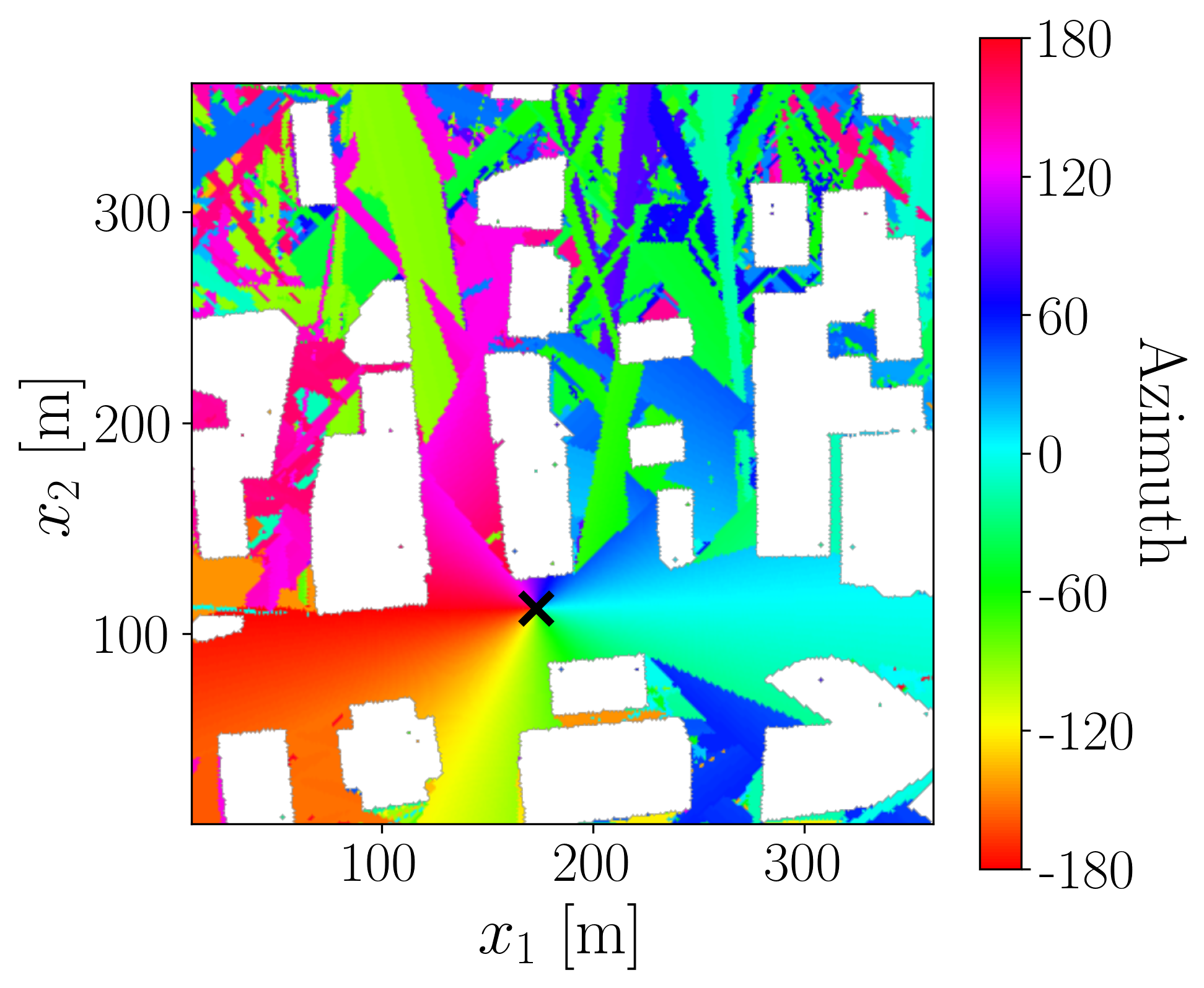}
        \vspace{-0.08in}
        \caption{\hspace*{-0.25in}}
        \label{subfig:beamforming_gt_az}
    \end{subfigure}
    \hfill
    \begin{subfigure}{0.55\columnwidth}
        \centering
        \includegraphics[trim={0.1in 0 0 0}, clip, keepaspectratio=true, width=1.85in]{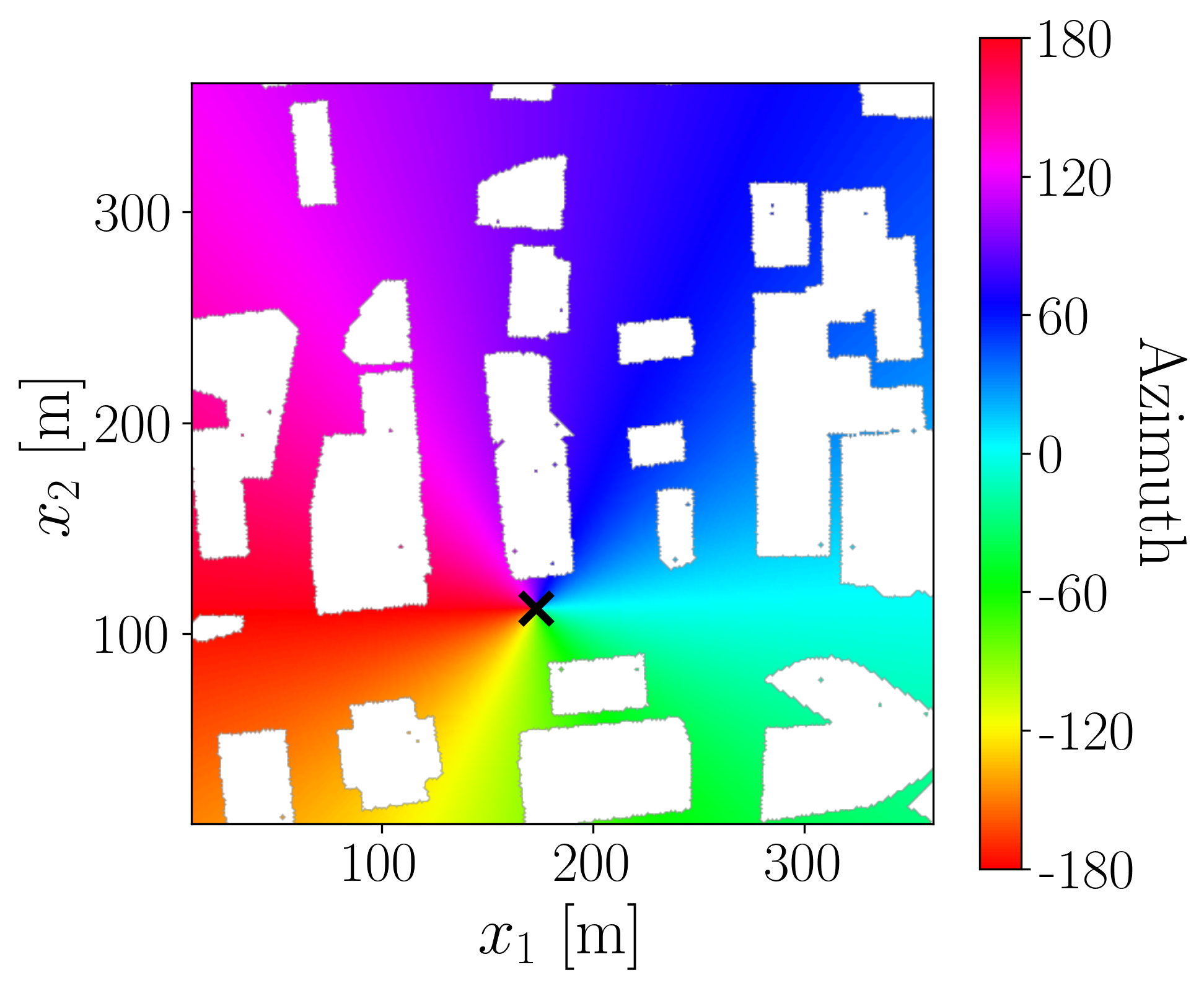}
        \vspace{-0.08in}
        \caption{\hspace*{0.1in}}
        \label{subfig:beamforming_pseudo_az}
    \end{subfigure}
    \vspace{-0.2in}
    \caption{
    Horizontal slice of the map of the urban region of interest in the beamforming experiments of Sec.~\ref{sec:experiments}.
    The layouts of the buildings are depicted as empty white regions, and the \glsentryshort{bs} position is marked by a black cross.
    The colored regions depict (a) the ground-truth and (b) the teacher model's azimuth angles of the strongest path at each device location.
    }
    \label{fig:bearmforming_setup}
    \vspace{-0.03in}
\end{figure}

\subsection{Optimal Tuning Parameter} \label{subsec:optimal_tuning}

\gls{cdr} addresses the minimization of the objective \eqref{eq:cdr_loss} using \gls{gd}.
Accordingly, at each iteration $t$, \gls{cdr} tackles the problem
\begin{equation}
\label{eq:opt_step}
    \htheta_{t+1} = \argmin_{\theta \in \Theta} L^{\CDR(\lambda_t)}_{\loss_t}(\theta),
\end{equation}
where 
$\loss_t(\theta| x, y) = \loss(\htheta_t | x, y) + \nabla_\theta \loss(\htheta_t | x, y)^\top (\theta - \htheta_t) + \frac{1}{2 \gamma_t} \norm{\theta - \htheta_t}^2$
is the local quadratic approximation of the loss function $\loss(\theta| x, y)$ around the current iterate $\htheta_t$ for learning rate $\gamma_t > 0$ \cite{simeone2022machine}.
The tuning vector $\lambda_t = [\lambda_{t, 1}, ..., \lambda_{t, K}]$ is chosen at each time $t$ to minimize the variance of the next parameter estimate $\htheta_{t+1}$ in \eqref{eq:opt_step}.
Following \cite[Prop.~2]{fisch2024stratified}, the ideal solution to this problem is obtained as
\begin{equation}
\label{eq:optimal_tuning_gd}
\lambda^{*}_{t, c} = \frac{
        \E_c\left[
            \left( \nabla \loss^f - \nabla \Loss^f_c \right)^\top
            \left( \nabla \loss - \nabla \Loss_c \right)
        \right]
    }{
        \left( 1 + \frac{n_c}{N_c} \right) \E_c\left[
            \Norm{\nabla \loss^f - \nabla \Loss^f_c}^2
        \right]
    },
\end{equation}
for $c \in \dset{K}$, where $\E_c[\cdot]$ is a shorthand for the context-dependent average $\E_{X, Y \sim P(X, Y | C=c)}[\cdot]$, and $\nabla \loss = \nabla_\theta \loss(\theta_t | X, Y)$ and $\nabla \loss^f = \nabla_\theta \loss(\theta_t | X, f(X))$ are the gradient and pseudo-gradient vectors at the current iterate $\theta_t$, with respective means $\nabla \Loss_c = \E_c[\nabla \loss]$ and $\nabla \Loss^f_c = \E_c[\nabla \loss^f]$.
Intuitively, the optimal parameter $\lambda^{*}_{t, c}$ measures the level of agreement between the centered gradients $(\nabla \loss - \nabla \Loss_c)$ and centered pseudo-gradients $(\nabla \loss^f - \nabla \Loss^f_c)$ within context $c$ via an averaged cosine similarity.
The ideal solution $\lambda^{*}_{t, c}$, however, depends on the true data distribution, which is unknown.
Therefore, \gls{cdr} estimates the expected values in \eqref{eq:optimal_tuning_gd} using the labeled data $\D_c$ as in \cite{fisch2024stratified}. 
Accordingly, the only additional complexity of \gls{cdr} as compared to \gls{dr} is the evaluation of the vector $\lambda^{*}_t$ with entries \eqref{eq:optimal_tuning_gd}, which has a complexity proportional to the number of trainable parameters, similar to the cost of a forward pass.

\begin{figure}
    \centering
    \includegraphics[keepaspectratio=true, width=2.8in]{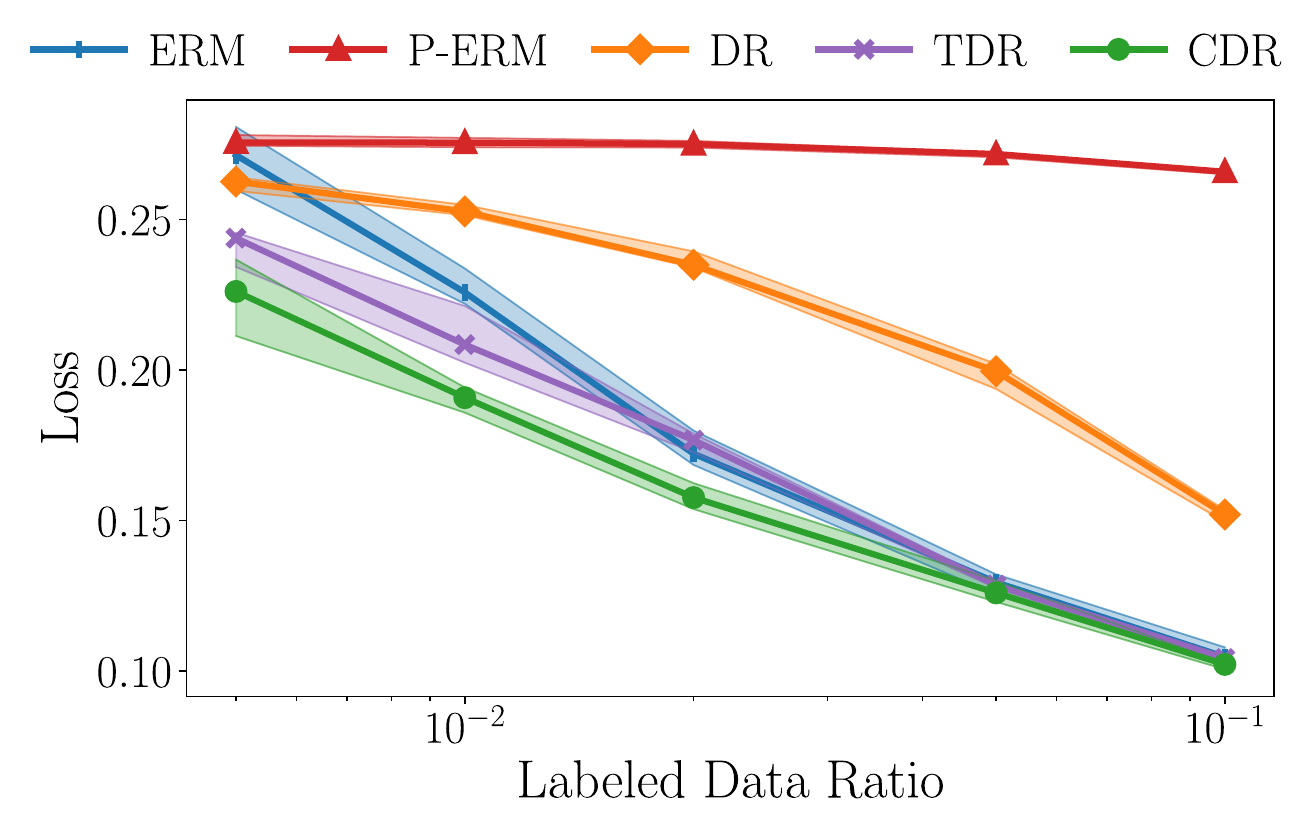}
    \vspace{-0.09in}
    \caption{
    Test loss as a function of the labeled data ratio $\rho$ for \glsentryshort{erm}, \glsentryshort{p-erm} \eqref{eq:pseudo_erm_loss}, \glsentryshort{dr} \cite{zhu2024doubly}, \glsentryshort{tdr}, and the proposed \glsentryshort{cdr} \eqref{eq:cdr_loss}.
    The solid lines represent the median of $10$ independent runs, and shaded areas indicate the range between the first and the third quartiles.
    }
    \label{fig:test_loss_vs_labeled_ratio}
    \vspace{-0.05in}
\end{figure}

\section{Experiments} \label{sec:experiments}

In this section, we study self-training for position-based downlink beamforming \cite{zeng2021toward, wu2021environment}.
The code used to carry the experiments is available at \cite{repoRuah2025CDR}.

\begin{figure*}
    \centering
    \begin{subfigure}{0.24\textwidth}
        \centering
        \includegraphics[keepaspectratio=true, width=1.7in]{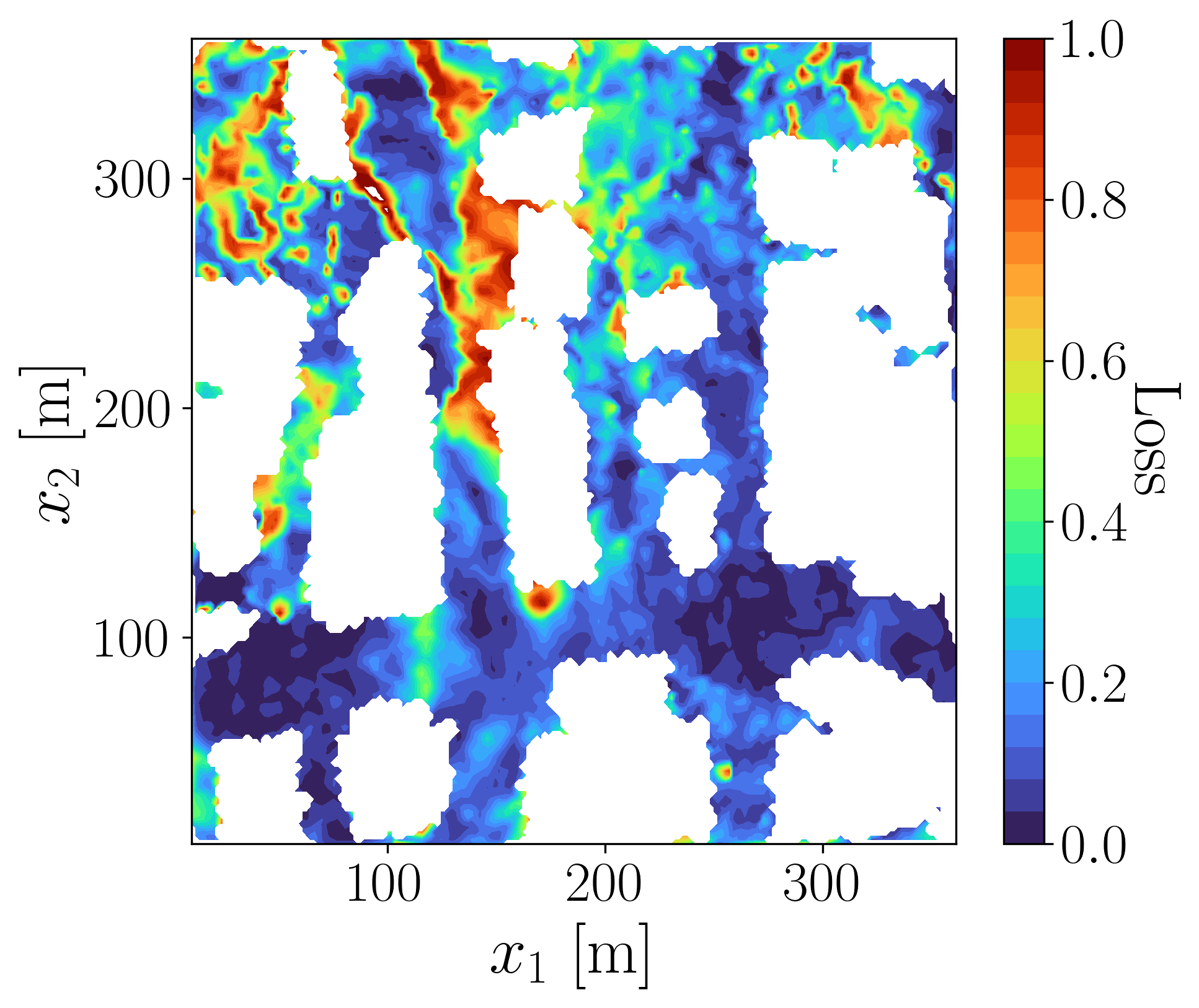}
        \vspace{-0.24in}
        \caption{\glsentryshort{erm}}
        \label{subfig:loss_map_erm}
    \end{subfigure}
    \hfill
    \begin{subfigure}{0.24\textwidth}
        \centering
        \includegraphics[keepaspectratio=true, width=1.7in]{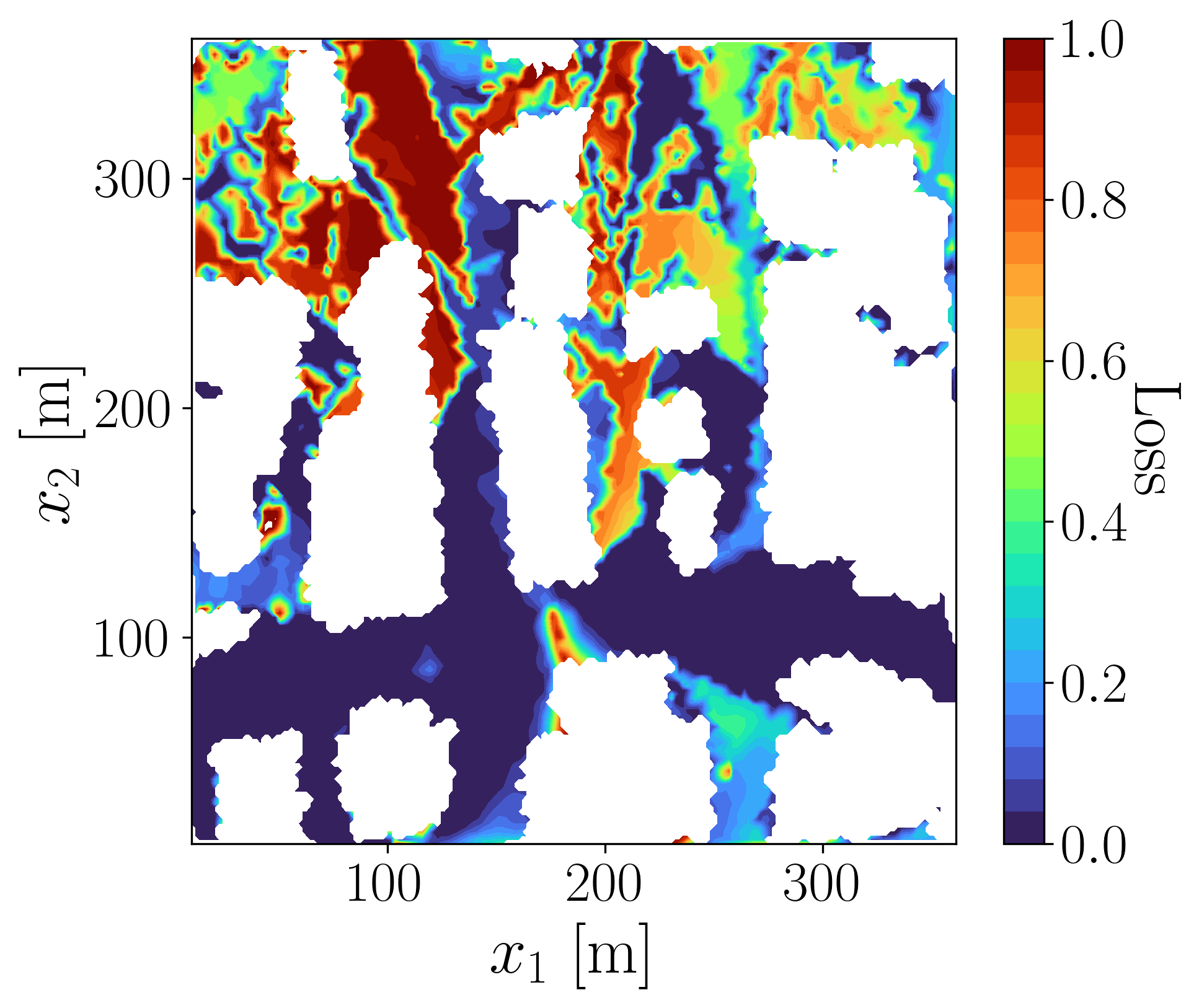}
        \vspace{-0.24in}
        \caption{\glsentryshort{p-erm}}
        \label{subfig:loss_map_pseudo_erm}
    \end{subfigure}
    \hfill
    \begin{subfigure}{0.24\textwidth}
        \centering
        \includegraphics[keepaspectratio=true, width=1.7in]{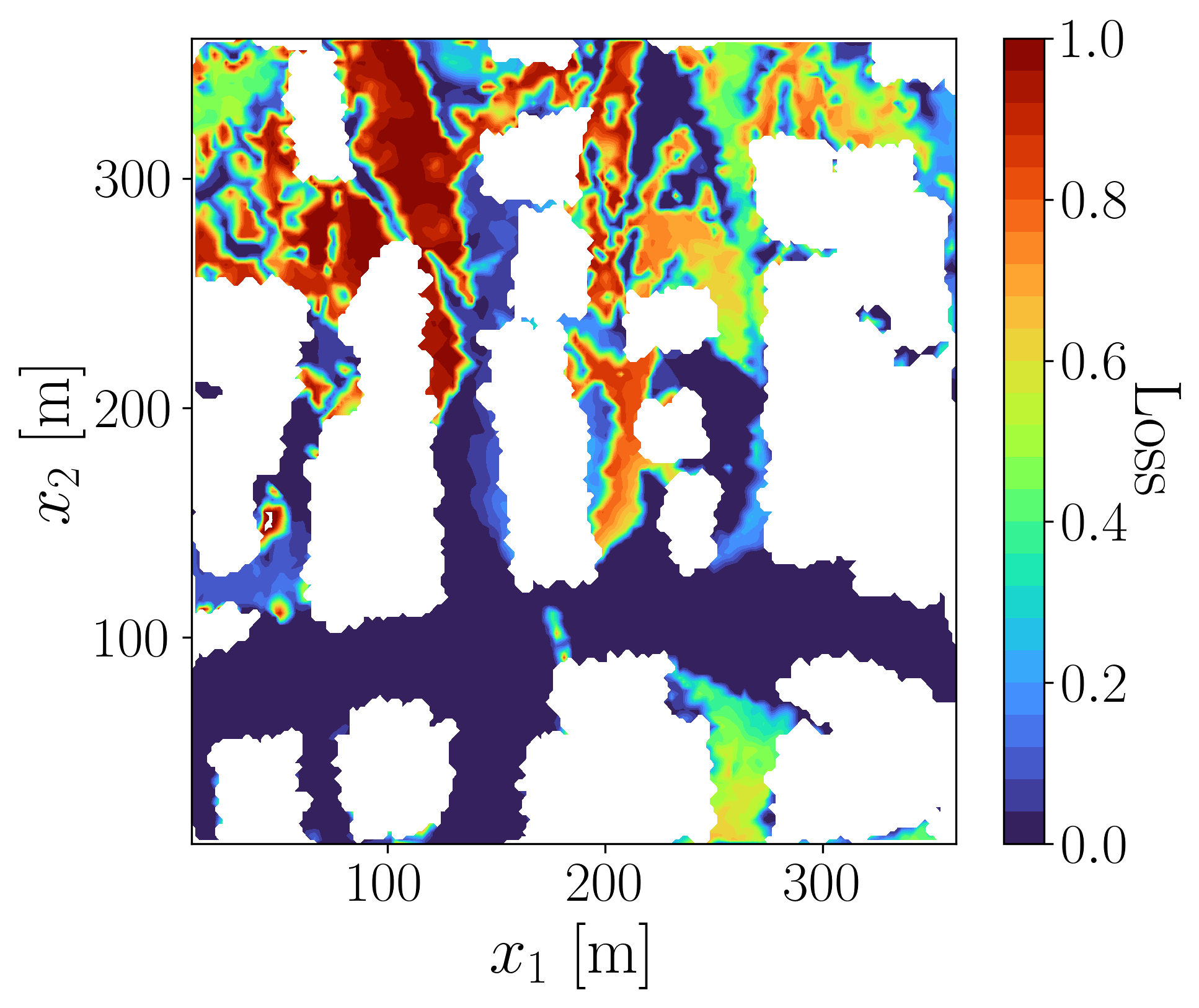}
        \vspace{-0.24in}
        \caption{\glsentryshort{dr}}
        \label{subfig:loss_map_dr}
    \end{subfigure}
    \hfill
    \begin{subfigure}{0.24\textwidth}
        \centering
        \includegraphics[keepaspectratio=true, width=1.7in]{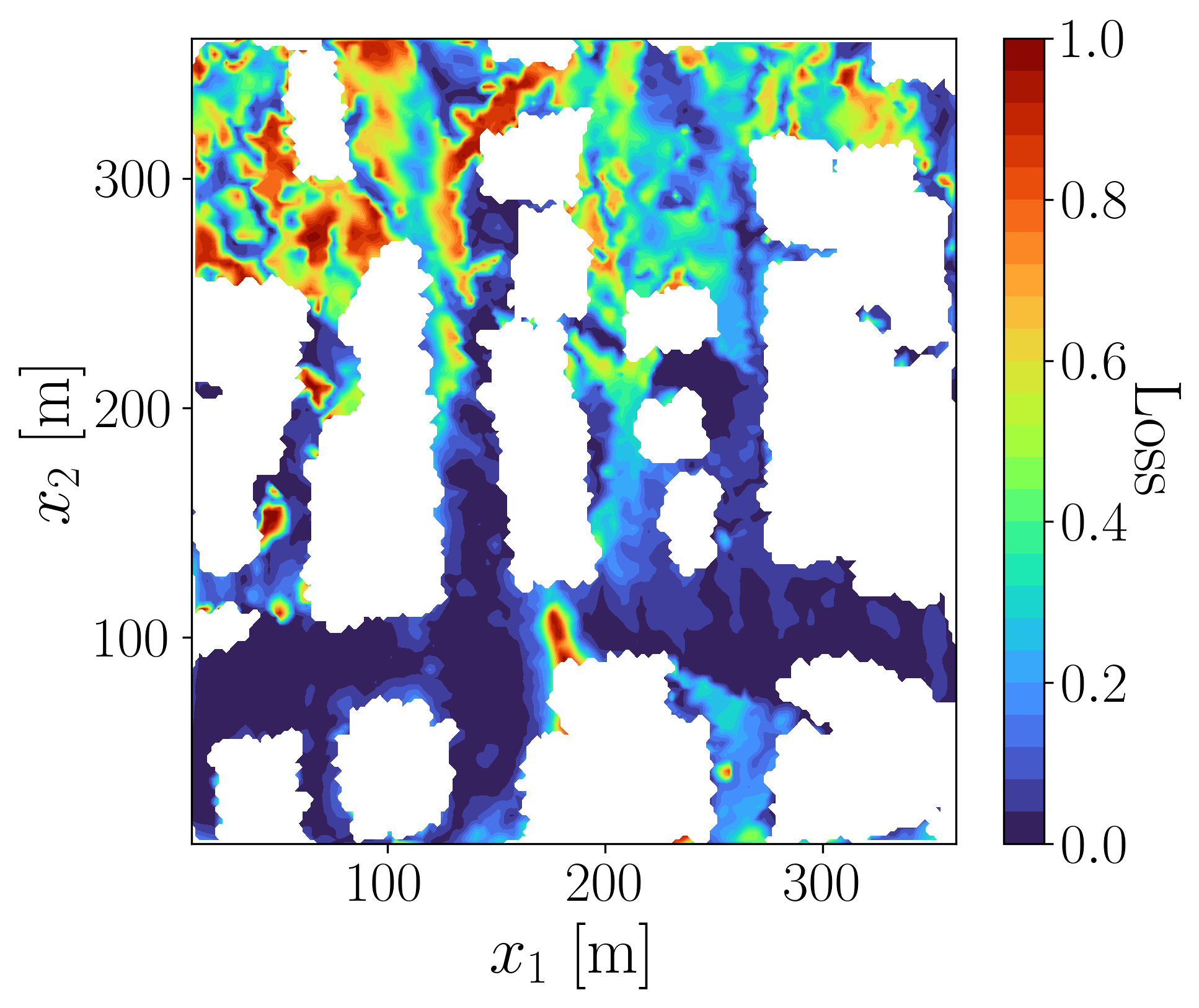}
        \vspace{-0.24in}
        \caption{\glsentryshort{tdr}}
        \label{subfig:loss_map_tdr}
    \end{subfigure}
    \begin{subfigure}{0.24\textwidth}
        \centering
        \includegraphics[keepaspectratio=true, width=1.7in]{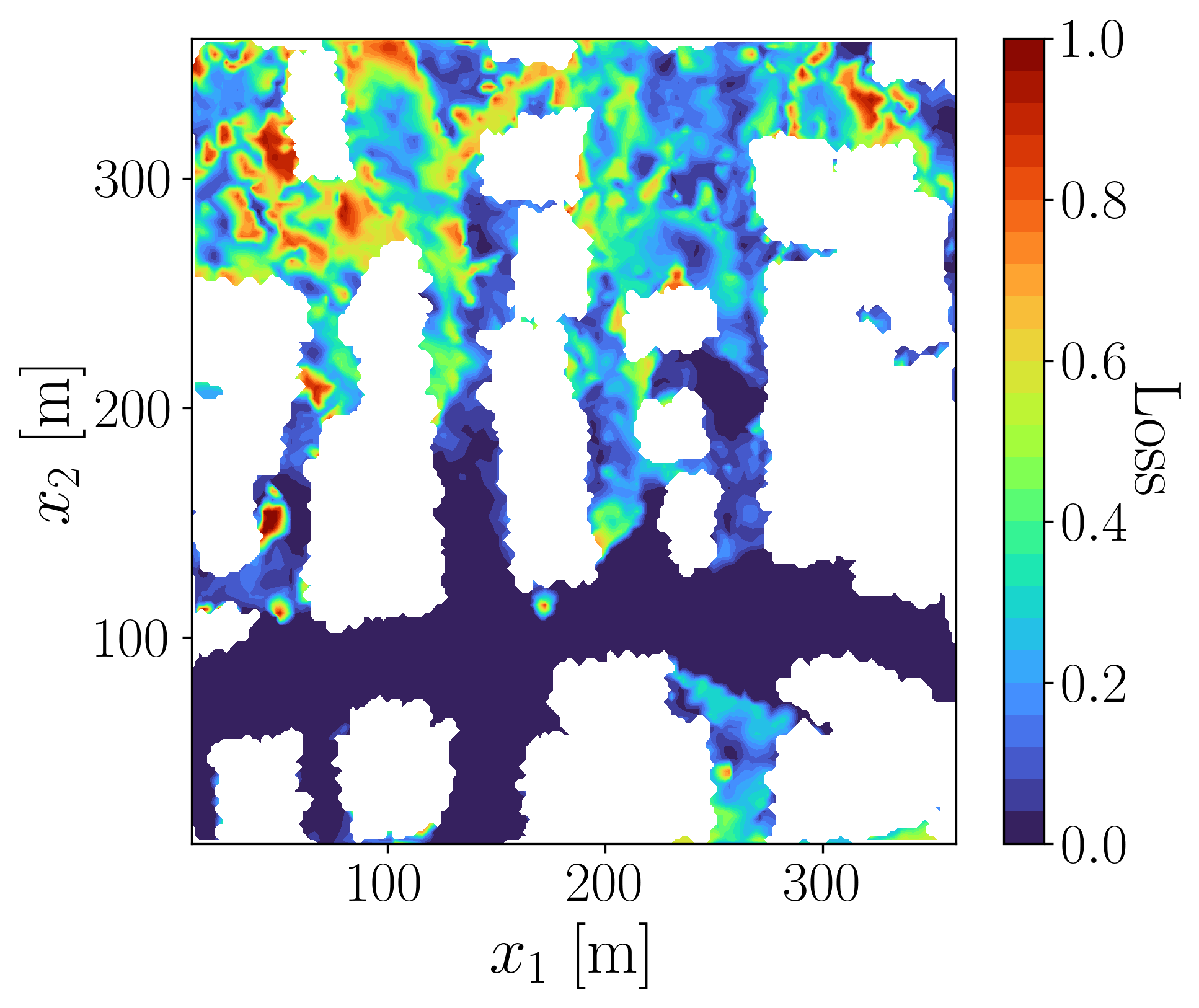}
        \vspace{-0.24in}
        \caption{\glsentryshort{cdr}}
        \label{subfig:loss_map_cdr}
    \end{subfigure}
    \begin{subfigure}{0.50\textwidth}
        \centering
        \includegraphics[keepaspectratio=true, width=3.35in]{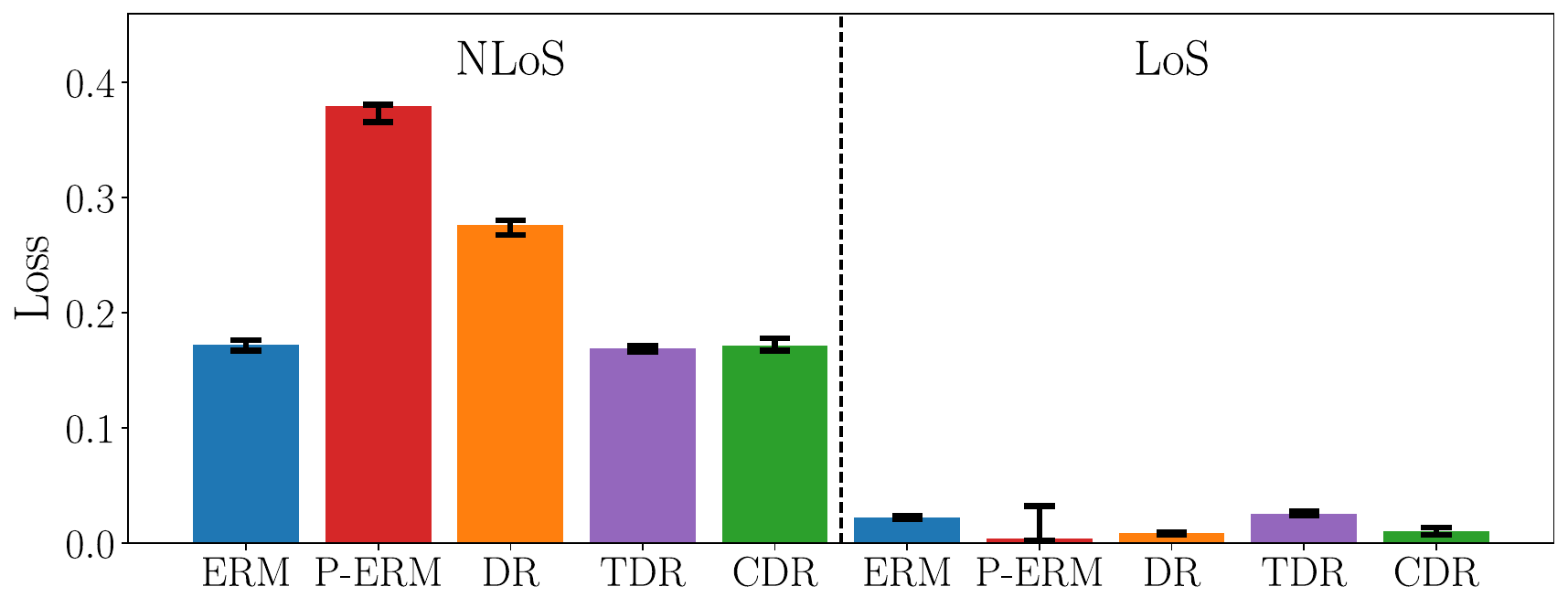}
        \caption{}
        \label{subfig:loss_per_context_barplot}
    \end{subfigure}
    \begin{subfigure}{0.24\textwidth}
        \centering
        \vspace{0.07in}
        \includegraphics[keepaspectratio=true, width=1.52in]{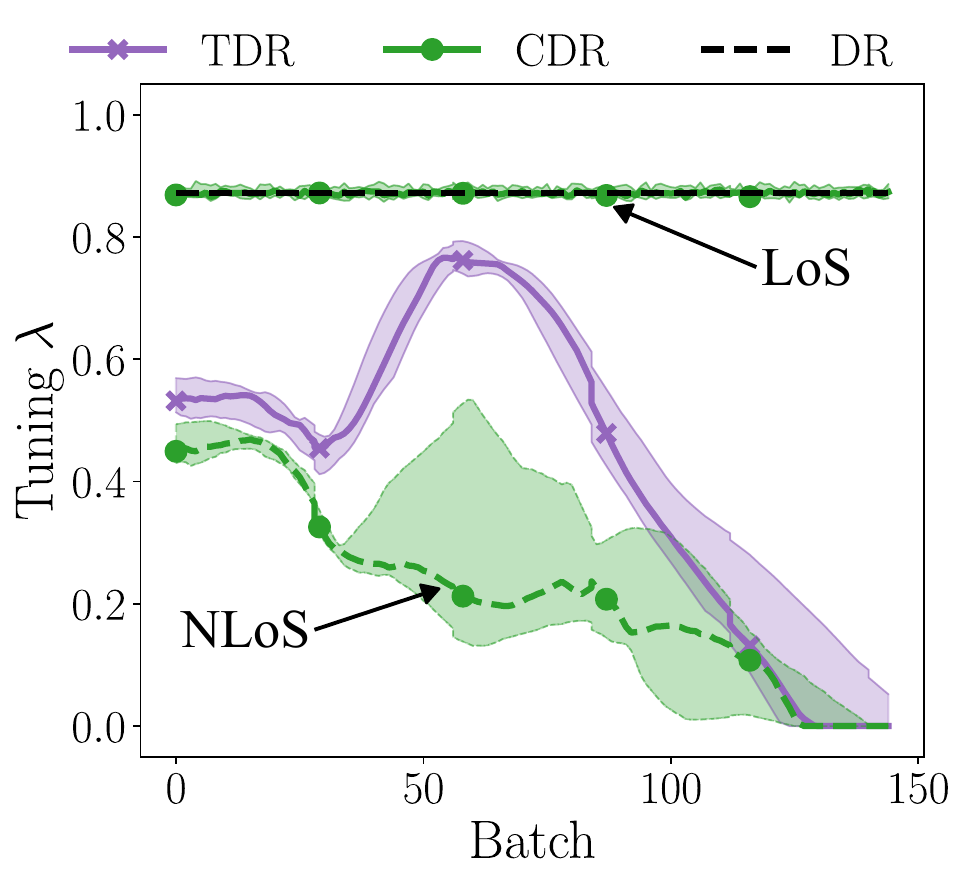}
        \vspace{-0.07in}
        \caption{}
        \label{subfig:tuning_parameter}
    \end{subfigure}
    \caption{
    Test loss maps for the (a) \glsentryshort{erm}, (b) \glsentryshort{p-erm} \eqref{eq:pseudo_erm_loss}, (c) \glsentryshort{dr} \cite{zhu2024doubly}, (d) \glsentryshort{tdr}, and (e) \glsentryshort{cdr} \eqref{eq:cdr_loss} with $n=150$ labeled data points.
    (f) Test loss per context.
    (g) Evolution of the computed tuning parameters $\lambda$ through training for \glsentryshort{tdr} (purple) and \glsentryshort{cdr} in \glsentryshort{nlos} (green dashed) and \glsentryshort{los} (green solid) conditions, as well as the effective tuning parameter $1 / (1 + n/N)$ assumed by \glsentryshort{dr} (black dashed).
    All levels correspond to the median of $10$ independent runs, and error bars in (f) and shaded areas in (g) indicate the range between the first and the third quartiles.
    }
    \label{fig:bearmforming_errors_distribution}
\end{figure*}

\inlinesection{Setup}
As depicted in Fig.~\ref{fig:bearmforming_setup}, we consider an outdoor urban setting with a single \gls{bs}, where propagation paths between the \gls{bs} and the devices are generated using ray-tracing as in \cite{zeng2021toward}.
As seen in Fig.~\ref{subfig:beamforming_gt_az} for the azimuth, each device position $X \in \R^3$ is associated with a ground-truth optimal azimuth and elevation \gls{aod} $Y = (Y^\az, Y^\el) \in \angleSet = [- \pi, \pi) \times [0, \pi]$ obtained by considering the propagation path with the highest received power.
Positions $X$ on the three dimensional space are uniformly sampled on the regions of the horizontal slice depicted in Fig.~\ref{fig:bearmforming_setup} that are not occupied by buildings.
The objective is to train a mapping $g_\theta: \R^3 \to \angleSet$ that outputs an estimate of the optimal \gls{aod} at each device location.

We produce a labeled training dataset of $N^\tr = 30000$ training pairs $\D^\tr = \{(x^\tr_i, y^\tr_i)\}_{i=1}^{N^\tr}$ and a labeled test dataset of $N^\te = 5975$ test pairs $\D^\te = \{(x^\te_i, y^\te_i)\}_{i=1}^{N^\te}$.
Furthermore, we augment each data point with a binary context variable $c_i \in \{0, 1\}$, where $c_i = 1$ indicates that the device at location $x_i$ is within \glsemph{los} of the \gls{bs}, and we set $c_i = 0$ otherwise, i.e., for \glsemph{nlos} coordinates.
The training dataset $\{(x^\tr_i, c^\tr_i, y^\tr_i)\}_{i=1}^{N^\tr}$ is randomly split into $n = \lceil \rho N^\tr \rceil$ labeled samples $\D = \{(x_i, c_i, y_i)\}_{i=1}^{n}$ and $N = N^\tr - n$ unlabeled samples $\tD = \{(\tx_i, \tc_i)\}_{i=1}^{N}$, where the labeled data ratio $\rho \in [0, 1]$ varies from $0.5\%$ to $10\%$.

The teacher model $f(x)$ disregards the geometry of the scene, producing pseudo-labels equal to the azimuth and elevation angles of the \gls{los} propagation path connecting the \gls{bs} and the device at location $x$.
As a result, as depicted in Fig.~\ref{subfig:beamforming_pseudo_az} for the azimuth, the teacher model is equal to the target angles in \gls{los} regions ($c = 1$), while presenting large discrepancies with the ground-truth target angles for \gls{nlos} conditions ($c = 0$).

\inlinesection{Implementation}
As in \cite{tancik2020fourier}, the trainable map $g_\theta$ is implemented as a feedforward neural network with an initial Fourier feature layer and two hidden layers with parameters $\theta$.
Writing the model's output as $g_\theta(x) = (\hat{y}^\az, \hat{y}^\el) \in \A$, the loss function is defined as the sum $\loss_\theta(x, y) = h(\hat{y}^\az, y^\az) + h(\hat{y}^\el, y^\el)$, where $y = (y^\az, y^\el)$ is the ground-truth target associated to position $x$, and $h(\hat{\varphi}, \varphi) = 1 - \cos(\hat{\varphi} - \varphi)$ is the log-loss for a von Mises model with independent azimuth and elevation angles with respective means $\hat{y}^\az$ and $\hat{y}^\el$ \cite{mardia2000directional}.
See Appendix~\ref{apx:implementation_details} for further implementation details.

\inlinesection{Benchmarks}
Apart from the \gls{p-erm} and \gls{dr} self-training procedures introduced in Sec.~\ref{sec:background}, we also consider the following benchmarks:
$1)$ \glsemph{erm}, which minimizes the standard loss $L_{\loss, \D}(\theta)$ limited to the labeled data; and $2)$ \glsemph{tdr} learning, which addresses the \gls{cdr} objective \eqref{eq:cdr_loss} with $\lambda_c = \lambda_0 \in [0, 1]$ for all contexts $c \in \dset{K}$, thus ignoring contextual information.

\gls{dr} is implemented by adopting the curriculum-based time-varying loss proposed in the original paper \cite{zhu2024doubly}, and a similar implementation is also used for \gls{tdr} and \gls{cdr} schemes, minimizing a time-varying loss of the form
\begin{equation}
\label{eq:cdr_loss_curriculum}
\begin{split}
L^{\CDR(\lambda)}_{\loss, \alpha_e}(\theta) = 
    &\sum_{c=1}^{K} \bigg\{
        \frac{\lambda_c N_c}{N} L_{\loss, \tD^f_c}(\theta) \\
        & + \alpha_e \left(
            \frac{n_c}{n} L_{\loss, \D_c}(\theta) -
            \frac{\lambda_c n_c}{n} L_{\loss, \D^f_c}(\theta)
        \right)
    \bigg\},
\end{split}
\end{equation}
with the linear schedule $\alpha_e = e/E$ over the total number of epochs $E$.

\inlinesection{Results}
Fig.~\ref{fig:test_loss_vs_labeled_ratio} presents the test loss $L_{\loss, \D^\te}(\htheta)$ for the estimates $\htheta$ described above as a function of the labeled data ratio.
Given that \gls{p-erm} and \gls{dr} do not have any mechanism to account for the accuracy of the teacher model $f$, they achieve a test loss larger than \gls{erm}, which neglects pseudo-labeled data.
In contrast, by selectively trusting the teacher model $f$ depending on the \gls{los}/\gls{nlos} context, the proposed \gls{cdr} outperforms all baselines, especially in the more challenging regime with limited labeled data.
For a labeled data ratio $\rho = 0.1$ ($n=300$ labeled samples), \gls{cdr} decreases the test loss of \gls{p-erm}, \gls{dr}, and \gls{tdr} by $31\%$, $24\%$, and $8\%$ respectively.

The variability of the test loss $\loss(\htheta | X, Y)$ as a function of the location $X$ is displayed on a two-dimensional map in Fig.~\ref{subfig:loss_map_erm}-\ref{subfig:loss_map_cdr} for $\rho = 0.005$, i.e., for $n=150$ labeled samples.
As emphasized in Fig.~\ref{subfig:loss_per_context_barplot}, \gls{p-erm} and \gls{dr} display near optimal performance in \gls{los} regions, where the teacher model $f$ is accurate, but they are outperformed by \gls{erm} in \gls{nlos} areas, where the teacher model $f$ is unreliable.
In contrast, \gls{tdr} closely aligns with \gls{erm}, since, as shown in Fig.~\ref{subfig:tuning_parameter}, the learned tuning parameter $\lambda_0$ converges to $0$ during training to accommodate the inaccuracies of $f$ on \gls{nlos} regions.
In comparison, as also shown in Fig.~\ref{subfig:tuning_parameter}, the proposed \gls{cdr} scheme accounts for the different accuracy levels of the teacher $f$ in \gls{los} and \gls{nlos} conditions, matching the performance of \gls{dr} in \gls{los} locations and of \gls{erm} in \gls{nlos} areas.

\section{Conclusion}

This work has proposed a novel \gls{cdr} semi-supervised learning scheme that adapts the importance of synthetically labeled samples to the accuracy of the teacher model at each given context via a tuning parameter vector.
An optimal choice of tuning parameter was derived as a function of the similarity between ground-truth and synthetically generated gradients.
Experiments on downlink beamforming empirically validated the advantage of \gls{cdr} over context-agnostic \gls{p-erm} and \gls{dr} self-training methodologies.
Future work may investigate the adaptation of \gls{cdr} to other successive convex approximation schemes \cite{marks1978general, razaviyayn2014parallel, liu2019stochastic}; the extension to continuous context variables using kernel methods \cite{hofmann2008kernel}; the use of the \gls{cdr} loss within other semi-supervised techniques \cite{sifaou2024semi}; and the design of computationally efficient approximations of the optimal tuning vector.

\bibliographystyle{IEEEtran}
\bibliography{refs}

\clearpage

\appendices

\section{Optimal Tuning for Local Convex Loss Approximation} \label{apx:optimal_tuning}

In this appendix, we detail the derivation of the optimal tuning parameter presented in Sec.~\ref{subsec:optimal_tuning}, starting from the more general case of an arbitrary choice of local convex approximation.
To avoid notational clutter, we use $\E_c[\cdot]$ and $\cov_c[\cdot]$ as shorthands for $\E_{X, Y \sim P(X,Y|C=c)}[\cdot]$ and $\cov_{X, Y \sim P(X,Y|C=c)}[\cdot]$.

\subsection{General Setting}

Denoting as $\loss_t(\theta | X, Y)$ a local convex approximation of the loss function around the current parameter iterate $\theta_t$, the choice of tuning parameter $\lambda^{*}_t = [\lambda^{*}_{t, 1}, ..., \lambda^{*}_{t, K}]$ that minimizes the asymptotic variance of the next parameter estimate $\htheta_{t+1} = \argmin_{\theta \in \Theta} L^{\CDR(\lambda_t)}_{\loss_t}(\theta))$ is obtained as a direct application of optimal power tuning in stratified \gls{ppi} \cite[Prop.~2]{fisch2024stratified}.
Accordingly, we have
\begin{equation}
\label{eq:optimal_tuning_general}
\begin{split}
\lambda^{*}_{t, c} = \frac{
        \tr_{\bar{H}_t}\left(
            C_{t, c}\left( \theta_{t+1} \right) +
            C_{t, c}\left( \theta_{t+1} \right)^\top
        \right)
    }{
        2 (1 + \frac{n_c}{N_c}) \tr_{\bar{H}_t}\left(
            \Vf_{t, c}\left( \theta_{t+1} \right)
        \right)
    },
\end{split}    
\end{equation}
with 
\begin{equation}
\tr_{\bar{H}_t}(A) = \tr\left( 
    \bar{H}_t\left( \theta_{t+1} \right)^{-1}
    A
    \bar{H}_t\left( \theta_{t+1} \right)^{-1}
\right)
\end{equation}
for an input matrix $A$, where $\theta_{t+1} = \argmin_{\theta \in \Theta} \E_{X, Y \sim P(X,Y)}[\loss_t(\theta | X, Y)]$ is the optimal next parameter iterate; 
\begin{equation}
    \Vf_{t, c}(\theta) = \cov_c[ \nabla_\theta \loss_t(\theta | X, f(X)) ]
\end{equation}
is the pseudo-loss gradient covariance; 
\begin{equation}
    C_{t, c}(\theta) = \cov_c\left[ \nabla_\theta \loss_t(\theta | X, Y), \nabla_\theta \loss_t(\theta | X, f(X)) \right]
\end{equation}
is the loss and pseudo-loss gradients cross-covariance; and $\bar{H}_t(\theta) = \E_{C \sim P(C)}[H_{t, C}(\theta)]$ is the average of the per-context Hessians 
\begin{equation}
    H_{t, c}(\theta) = \E_c[\nabla^{2}_{\theta} \loss_t(\theta | X, Y)].
\end{equation}

\subsection{Gradient Descent Quadratic Approximation}

We now simplify the optimal tuning formula in \eqref{eq:optimal_tuning_general} for the case of \gls{gd} optimization, which is implemented using a quadratic local approximation of the form
\begin{multline}
\label{eq:convex_approx_gd_apx}
\loss_t(\theta| x, y) = \\
    \loss(\theta_t | x, y) + \nabla_\theta \loss(\theta_t | x, y)^\top (\theta - \theta_t) + \frac{1}{2 \gamma_t} \norm{\theta - \theta_t}^2,
\end{multline}
where $\gamma_t > 0$ is the learning rate, and where $\norm{\cdot}$ denotes the Euclidean norm.

We define as $\nabla \loss = \nabla_\theta \loss(\theta_t | X, Y)$ and $\nabla \loss^f = \nabla_\theta \loss(\theta_t | X, f(X))$ the gradient and pseudo-gradient vectors at the current parameter estimate $\theta_t$, and denote their global and context-dependent means as $\nabla \Loss = \E_{X, Y \sim P(X,Y)}[\nabla \loss]$, $\nabla \Loss_c = \E_c[\nabla \loss]$, and $\nabla \Loss^f_c = \E_c[\nabla \loss^f]$.
From the \gls{gd} local approximation \eqref{eq:convex_approx_gd_apx}, we have $\theta_{t+1} = \theta_t - \gamma_t \nabla \Loss$; $\nabla_\theta \loss_t(\theta_{t+1} | X, Y) = \nabla \loss - \nabla \Loss$; $\bar{H}_t\left( \theta_{t+1} \right) = \gamma_t^{-1} I$; $\E_c[\nabla_\theta \loss_t(\theta_{t+1} | X, Y)] = \nabla \Loss_c - \nabla \Loss$; and $\E_c[\nabla_\theta \loss_t(\theta_{t+1} | X, f(X))] = \nabla \Loss^f_c - \nabla \Loss$, where $I$ denotes the identity matrix.
Accordingly, the cross-covariance term in \eqref{eq:optimal_tuning_general} can be expressed as
\begin{equation}
    C_{t, c}\left( \theta_{t+1} \right) = \E_c\left[
        \left( \nabla \loss - \nabla \Loss_c \right) 
        \left( \nabla \loss^f - \nabla \Loss^f_c \right)^\top
    \right].
\end{equation}
Using the invariance of the trace operator to cyclic permutations, we can compute the denominator in \eqref{eq:optimal_tuning_general} as 
\begin{equation}
\begin{split}
&\tr_{\bar{H}_t}\left(
    C_{t, c}\left( \theta_{t+1} \right)
\right) \\
    &\quad = \tr \left(
        \gamma_t I 
        \E_c\left[
            \left( \nabla \loss - \nabla \Loss_c \right)
            \left( \nabla \loss^f - \nabla \Loss^f_c \right)^\top
        \right]
        \gamma_t I 
    \right) \\
    &\quad = \gamma_t^2 \E_c\left[
        \tr\left(
            \left( \nabla \loss - \nabla \Loss_c \right)
            \left( \nabla \loss^f - \nabla \Loss^f_c \right)^\top
        \right)
    \right] \\
    &\quad = \gamma_t^2 \E_c\left[
        \left( \nabla \loss^f - \nabla \Loss^f_c \right)^\top
        \left( \nabla \loss - \nabla \Loss_c \right)
    \right].
\end{split}
\end{equation}
Similarly, we have
\begin{multline}
\tr_{\bar{H}_t}\left(
    \Vf_{t, c}\left( \theta_{t+1} \right)
\right) = \\
    \gamma_t^2 \E_c\left[
        \left( \nabla \loss^f - \nabla \Loss^f_c \right)^\top
        \left( \nabla \loss^f - \nabla \Loss^f_c \right)
    \right].
\end{multline}
All in all, the optimal tuning parameter for gradient descent can be expressed as
\begin{equation}
\begin{split}
\lambda^{*}_{t, c} = \frac{
        \E_c\left[
            \left( \nabla \loss^f - \nabla \Loss^f_c \right)^\top
            \left( \nabla \loss - \nabla \Loss_c \right)
        \right]
    }{
        \left( 1 + \frac{n_c}{N_c} \right) \E_c\left[
            \Norm{\nabla \loss^f - \nabla \Loss^f_c}^2
        \right]
    }.
\end{split}    
\end{equation}

Since the ideal solution $\lambda^{*}_{t, c}$ depends on the unknown data distribution $P(X, Y | C=c)$, it is instead estimated using the labeled data $\D_c$ as
\begin{multline}
\hat{\lambda}_{t, c} = \\
    \frac{
        \frac{1}{n_c - 1}
        \sum_{i=1}^{n_c}
            \left( \nabla \loss^f_{c, i} - \nabla \widehat{\Loss}^f_c \right)^\top
            \left( \nabla \loss_{c, i} - \nabla \widehat{\Loss}_c \right)
    }{
        \left( 1 + \frac{n_c}{N_c} \right) 
        \frac{1}{n_c - 1}
        \sum_{i=1}^{n_c}
            \Norm{\nabla \loss^f_{c, i} - \nabla \widehat{\Loss}^f_c}^2
    },
\end{multline}
where $\nabla \loss_{c, i} = \nabla_\theta \loss(\theta_t | x^c_i, y^c_i)$, $\nabla \loss^f_{c, i} = \nabla_\theta \loss(\theta_t | x^c_i, f(x^c_i))$, $\nabla \widehat{\Loss}_c = \frac{1}{n_c} \sum_{i=1}^{n_c} \nabla \loss_{c, i}$, and $\nabla \widehat{\Loss}^f_c = \frac{1}{n_c} \sum_{i=1}^{n_c} \nabla \loss^f_{c, i}$.

\section{Implementation Details} \label{apx:implementation_details}

The trainable map $g_\theta$ in Sec.~\ref{sec:experiments} is implemented as a feedforward neural network with an initial Fourier feature layer \cite{tancik2020fourier} and two hidden layers composed of $128$ and $64$ neurons respectively with ReLU activations.
The Fourier feature layer is implemented by concatenating the $m=20$ features $p(v) = \{ (\cos(2\pi \sigma^{j/m} v), \sin(2\pi \sigma^{j/m} v) \}_{j=0}^{m-1}$ for each coordinate $v$ of the input position $x \in \R^3$, with $\sigma = 20$.
All models are trained until convergence over $E=100$ epochs, with the exception of the ERM model that is trained over $E=1000$ epochs to compensate for the reduced number of available training samples.
Parameter optimization uses an Adam optimizer with $\beta_1=0.9$, $\beta_2=0.999$ and constant learning rate $\gamma = 5 \times 10^{-4}$ \cite{kingma2014adam}.

\end{document}